\documentclass[12pt]{article}

\catcode`\@=11

\global\arraycolsep=2pt
\usepackage{amsbsy,amssymb,latexsym,amsfonts,amsmath}
\usepackage{graphicx,color}

\begin{document}
\begin{flushright}
\parbox{4.2cm}
{RUP-17-3}
\end{flushright}

\vspace*{0.7cm}

\begin{center}
{ \Large Can we change $c$ in four-dimensional CFTs by exactly marginal deformations?}
\vspace*{1.5cm}\\
{Yu Nakayama}
\end{center}
\vspace*{1.0cm}
\begin{center}

Department of Physics, Rikkyo University, Toshima, Tokyo 171-8501, Japan

\vspace{3.8cm}
\end{center}

\begin{abstract}
There is no known obstructions, but we have not been aware of any concrete examples, either. The Wess-Zumino consistency condition for the conformal anomaly says that $a$ cannot change but does not say anything about $c$. In supersymmetric models,  both $a$ and $c$ are determined from the triangle t'Hooft anomalies and the unitarity demands that both must be fixed, so the unitary supersymmetric conformal field theories do not admit such a possibility.
 Given this field theory situation, we construct an effective AdS/CFT model without supersymmetry in which $c$ changes under exactly marginal deformations. 

\end{abstract}

\thispagestyle{empty} 

\setcounter{page}{0}

\newpage

\section{Introduction}
Counting degrees of freedom in quantum field theories is the first step for the classification of the entire landscape of our theory space. It is long known that in two-dimensional conformal field theories, the central charge $c$ of the Virasoro algebra 
\begin{align}
[L_m,L_n] = (m-n) L_{m+n} + \frac{c}{12}(m^3-m)\delta_{m+n,0}
\end{align}
plays such a role. The central charge $c$ may also be regarded as a coefficient of the Weyl anomaly: the trace of the energy-momentum tensor does not vanish
\begin{align}
T^i_{\ i} = \frac{c}{24\pi} R \ 
\end{align}
in the curved background.
In unitary conformal field theories with a discrete conformal spectrum, the central charge $c$ monotonically decreases along the renormalization group flow and it stays constant along the exactly marginal deformations \cite{Zamolodchikov:1986gt}. Because of this property, the central charge $c$ is a good candidate for representing the degrees of freedom in conformal field theories, and indeed, the first thing to know in the classification of the conformal field theories is the central charge $c$.

In four-dimensions, it took more than two decades to completely settle the issue \cite{Cardy:1988cwa}\cite{Komargodski:2011vj}. In four-dimensional conformal field theories, there are two independent terms in the Weyl anomaly: schematically we have 
\begin{align}
T^i_{\ i} = -a \mathrm{Euler} + c \mathrm{Weyl}^2
\end{align}
 and it eventually turned out that one of them called ``$a$" plays the similar role as the central charge $c$ in two-dimensions. Historically, however, it even took some years that the other combinations of $a$ and $c$ are not good candidates for representing the degrees of freedom. The first counterexample in which $c$ increases along the renormalization group flow  appeared in the context of the supersymmetric gauge theories, in which non-perturbative determination of $a$ and $c$ is possible \cite{Anselmi:1997am}.

The Weyl anomaly coefficient $a$ has a nice property that it cannot depend on the exactly marginal deformations of any conformal field theories in four-dimensions. Thus, from the beginning it is a good candidate for representing the degrees of freedom as the central charge $c$ in two-dimensions. However, the question has been still open if we can actually change $c$ along exactly marginal deformations. If it does, then it cannot be a candidate for representing the degrees of freedom.  In fact, there is no known obstructions, but we have not been aware of any concrete examples, either. The Wess-Zumino consistency condition for the conformal anomaly says that $a$ cannot change but does not say anything about $c$ \cite{Osborn:1991gm}. 

The aim of this paper is to study a possibility to change $c$ along exactly marginal deformations. We eventually construct such a model in the effective AdS/CFT correspondence. Therefore, as an effective theory of gravity, we open up a possibility for a model with varying $c$ along exactly marginal deformations. Then we may further ask why it is more difficult to find such a situation in quantum field theories, and this is another aim of this paper.

To conclude the introduction, we should mention that the proof that $a$ is monotonically decreasing along the renormalization group flow appeared in \cite{Komargodski:2011vj}, and from that viewpoint alone, there remains a  less direct motivation to study the properties of $c$. Nevertheless, we think it is an interesting tension between ``what is allowed must happen in physics" and ``we do not know any concrete examples". Even beyond the theoretical curiosity, studying properties of the energy-momentum tensor is certainly one of the most fundamental aspects of conformal field theories (e.g. \cite{Osborn:1993cr}\cite{Erdmenger:1996yc}), and it is important to understand them better, in particular without supersymmetry where we have less control over them.

The organization of the paper is as follows. We are going to set up a debate if  we can change $c$ in four-dimensional CFTs by exactly marginal deformations. In section 2, we present the field theory argument with a pro and a con. In section 3, we continue the argument from the holographic perspective with a pro and a con. In section 4, we conclude the debate with further discussions.

\section{Field theory argument}

\subsection{Pro}

{\bf Statement:} 

{\it There must be a model in which $c$ changes under exactly marginal deformations.  As we will show, the most powerful constraint on the Weyl anomaly coming from the Wess-Zumino consistency conditions for the local renormalization group transformation allows such a  possibility, so from the totalitarian  principle of theoretical physics \cite{Gell-Mann:1956iqa} ``Everything not forbidden is compulsory" dictates such a model should exist.}

We first show that there is a non-perturbative field theory argument that in four-dimensional conformal field theories, the Weyl anomaly coefficient $a$ cannot depend on exactly marginal coupling constants while $c$ may depend on them. The argument was given in \cite{Osborn:1991gm} as a Wess-Zumino consistency condition for the Weyl transformation with the space-dependent coupling constant. This is also known as a local renormalization group analysis, and we will see that not only space-time dependent metric but also space-time dependent coupling constant plays a crucial role.

Let $\lambda(x)$ be a space-time dependent exactly marginal coupling constant of a conformal field theory, which means that if we put the theory on the flat Minkowski space-time $g_{ij}(x) = \eta_{ij}$ with the space-time independent coupling constant $\lambda(x) = \lambda$, the theory is conformal invariant (for a certain range of $\lambda$, which we will not bother in the following discussions). 
Even though the theory is conformal invariant on the flat Minkowski space with a space-time independent coupling constant, the variation of the partition functional under the Weyl transformation in the curved background with the space-time dependent coupling constant may not vanish due to the Weyl anomaly: under the infinitesimal Weyl variation of the metric: $\delta_\sigma g_{ij}(x) = 2\sigma(x) g_{ij}(x)$, we have a variation of the partition functional $Z[g_{ij},g]$
\begin{align}
\delta_{\sigma} \log Z[g_{ij}(x),\lambda(x)] = \int d^4x \sqrt{|g|} 2\sigma(x) \left(a(\lambda) \mathrm{Euler} - c(\lambda) \mathrm{Weyl}^2 + \cdots \right) \ ,
\end{align}
where $\cdots$ represents the Weyl anomaly coming from the space-time dependent coupling constant such as $\Box \lambda(x) \Box \lambda(x)$, which, for now, we are not interested in.

The Weyl anomaly must satisfy an integrability condition. To see this, we note that the Weyl variation is Abelian 
\begin{align}
\delta_{\sigma} \delta_{\tilde{\sigma}} \log Z[g_{ij}(x), \lambda(x)] -  \delta_{\tilde{\sigma}} \delta_{{\sigma}} \log Z[g_{ij}(x), \lambda(x)] = 0 \ , \label{Weylc}
\end{align}
and this gives a non-trivial constraint on the form of the Weyl anomaly. For example, the pure $R^2$ term in the Weyl anomaly (i.e. $T^{i}_{\ i} = b R^2$) is not allowed  just from this integrability condition \cite{Bonora:1985cq}:
\begin{align}
\delta_\sigma \delta_{\tilde{\sigma}} \log Z[g_{ij}(x), \lambda(x)] -  \delta_{\tilde{\sigma}} \delta_{{\sigma}} \log Z[g_{ij}(x), \lambda(x)] = \int d^4x \sqrt{|g|}b R \left(  \tilde{\sigma} \Box \sigma - \sigma \Box \tilde{\sigma}\right) \ .
\end{align}

Let us see what this constraint dictates about the properties of $a(\lambda)$ and $c(\lambda)$.
First of all, we note that the Weyl squared term is invariant under the Weyl transformation $\delta_{\sigma} \left( \sqrt{|g|} \mathrm{Weyl}^2 \right) = 0$, so  there is no constraint on the coefficient $c(\lambda)$ from \eqref{Weylc}. In particular, there is no obstruction for $c$ to depend on the (exactly marginal) coupling constant $\lambda$.

On the other hand, the Weyl variation of the Euler density is non-trivial: under the infinitesimal Weyl variation (i.e. $\delta g_{ij} = 2\sigma g_{ij}$), it transforms as 
\begin{align}
\delta_{\sigma} (\sqrt{|g|} \mathrm{Euler}) = 8\sqrt{|g|}\left(R_{ij} -\frac{R}{2}g_{ij} \right) D^i\partial^j \sigma \ .
\end{align}
Therefore the Wess-Zumino consistency condition demands
\begin{align}
[\delta_{\tilde{\sigma}}, \delta_{\sigma}]  \log Z[g_{ij}(x), \lambda(x)] = \int d^4x \sqrt{|g|} \left(8a(\lambda) \left(R_{ij} -\frac{R}{2}g_{ij} \right) (\sigma D^i\partial^j \tilde{\sigma} - \tilde{\sigma} D^i \partial^j \sigma) + \cdots \right.
\end{align}
However, when $a(\lambda)$ is a function of space-time dependent coupling constants such that $a(\lambda)$ is a non-trivial function of $x^i$, one cannot do the integration by part and the right hand side does not vanish while the left hand side should, leading to a contradiction. Only when $a$ is a genuine constant, one may perform the integration by part with the Bianchi identity to show that it is consistent.
 This means that the Weyl anomaly coefficient $a$ cannot depend on the exactly marginal coupling constant. We emphasize that this conclusion only comes from the consideration of space-time dependent coupling constant with local renormalization group analysis.

In order to make this argument complete, we have to further show that the omitted terms coming from the space-time dependent coupling constants do not give rise to the term $\left(R_{ij} -\frac{R}{2}g_{ij} \right) (\sigma D^i\partial^j \tilde{\sigma} - \tilde{\sigma} D^i \partial^j \sigma)$ for a possible cancellation, and we can easily see this is indeed the case. It is essentially because the other terms already contain the derivatives of $\lambda(x)$ and the Weyl variation cannot generate the Einstein tensor once the beta function for the coupling constant $\lambda$ vanishes. For example, the variation of $\int d^4x \sqrt{|g|} \sigma(x) \Box \lambda(x) \Box \lambda(x)$ is $\int d^4x \sqrt{|g|} \sigma(x) \partial^i \tilde{\sigma} \partial_i \lambda \Box \lambda(x)$ and cannot be used to cancel the above variation from the Euler density whose variation contains the Einstein tensor.\footnote{When $\lambda$ has a non-vanishing beta function, one may be able to cancel it from the terms such as $\int d^4x \sqrt{|g|} R_{ij}\partial^i \lambda(x) \partial^j \lambda(x)$, and this observation is a starting point to discuss $a$-theorem from the local renormalization group analysis, but it is another story.} The more details can be found in \cite{Osborn:1991gm}, but we do not need it for the rest of our discussions because our focus is $c$ rather than $a$.

We therefore conclude that from the totalitarian principle of physics there must exist a model in which $c$ changes under exactly marginal deformations. 

\subsection{Con}

{\bf Statement:}
 
{\it There are no known examples in which $c$ changes under exactly marginal deformations. First of all, there is a fine-tuning problem to have exactly marginal deformations, and if we overcome this difficulty with the supersymmetry, then the structure of the supersymmetric field theory together with unitarity demands that $c$ cannot change under exactly marginal deformations. Thus the situations presented in section 2.1 are just a rice-cake in the picture.}\footnote{A Japanese proverb corresponding to a pie in the sky or just wishful thinking.}

The conclusion made in section 2.1 is that from the kinematic structure of the effective action alone, there is  no obstruction that the Weyl anomaly coefficient $c$, unlike $a$, depends on exactly marginal deformations of conformal field theories. As far as we know, however, there have been no concrete field theory examples that show this property.

The question whether we can change the Weyl anomaly coefficient $c$ in four-dimensional CFTs by exactly marginal deformations, first of all, relies on the possibility that we have exactly marginal deformations. This itself is highly non-trivial. Typically the beta functions do not vanish and we need either symmetry reasoning or controlling parameters with fine-tunings to ensure it. Otherwise it would be quite accidental.

Only after setting up concrete ways to realize exactly marginal deformations, we should look for the change of $c$ under the exactly marginal deformations. 
The best known strategy to obtain exactly marginal deformations in four-dimensional conformal field theories is to impose the supersymmetry. The non-renormalization theorem  allows a possibility to have exactly marginal deformations in supersymmetric field theories. 
Then the second natural question to be asked is whether we can change $c$ in four-dimensional superconformal field theories by exactly marginal deformations.

Let us look at the Wess-Zumino consistency conditions for the super Weyl anomaly \cite{Grosse:2007au}\cite{Auzzi:2015yia}. The term containing $c$ is uplifted to
\begin{align}
\delta_{\sigma}\log Z= \int d^4x d^2\theta \left(\Sigma \kappa(\lambda) W_{\alpha\beta\gamma} W^{\alpha\beta\gamma} + \mathrm{c.c} \right) \ ,
\end{align}
where $\Sigma$ is a chiral super Weyl variation, and $\kappa(\lambda) = a-c(\lambda)$ with $\lambda$ being a chiral background superfield (corresponding to an exactly marginal deformation), and $W_{\alpha\beta\gamma}$ is the Weyl superfield. Note that we already know that $a$ cannot depend on the exactly marginal deformations.

This alone does not give us a constraint on $\kappa(\lambda)$ from the Wess-Zumino consistency condition because the Weyl superfield squared is super Weyl invariant. However, if we expand it in terms of the component, the Weyl superfield contains a field strength that couples with the  superconformal R-current, and together with the term containing $a$ Weyl anomaly coefficient, we may relate the R-current t'Hooft anomaly to $a$ and $c$:
\begin{align}
a &= \frac{9}{16 (8\pi)^2}\left(3\mathrm{Tr} R^3 - \mathrm{Tr} R  \right)\cr
c &= \frac{9}{16 (8\pi)^2}\left(3\mathrm{Tr} R^3 - \frac{5}{3}\mathrm{Tr} R \right) \ .
\end{align}
Here $\mathrm{Tr}$ means the t'Hooft triangle anomaly, or more precisely, in the conformal field theory language, it is computed by a particular three-point functions among three R-currents or one R-current and two energy-momentum tensor. This relation tells that in order to change $c$ anomaly along the exactly marginal deformations, we need to change the R-current t'Hooft anomaly while keeping the particular combination $a$ fixed. Furthermore, there is a so-called $a$-maximization procedure \cite{Intriligator:2003jj} to determine the superconformal $R$ symmetry: the superconformal R-symmetry is the R-symmetry such that the above expression for $a$ is maximized.

This makes the situation already difficult but we admit that kinematically it is still not impossible. For example, let us consider the case with $U(1)^3$ symmetry whose charges are denoted by $x$,$y$,$z$. Take a particular R-charge constraint $x-y+z=1$, and the trial $a$-function\footnote{This example is just for illustration and does not represent a unitary quantum field theory, if any. See below.}
\begin{align}
a &= -3\left((x-y+z)((x-y)^2+z^2) + (x-y+z)^2(x+y) \right) + 3(x+y) \cr
  & = -3 \left((1-z)^2 + z^2 \right) 
\end{align}
and we see that the $a$-maximization determines $a$ at $z=\frac{1}{2}$, but the theory still has one parameter moduli, say in $y$ direction (with $x=\frac{1}{2}+y)$. On the other hand, the same t'Hooft anomaly says
\begin{align}
c &= -3\left((x-y+z)((x-y)^2+z^2) + (x-y+z)^2(x+y) \right) + 5(x+y) \cr
  & = -3 \left((1-z)^2 + z^2 \right) +2(1+2y-z) 
\end{align}
and it depends on the moduli $y$ that was not fixed by the $a$-maximization.

This model or any similar kind, however, does not arise in unitary supersymmetric field theories for the following reason. The unitarity demands that two-point functions of the conserved current operators must be positive definite, but the superconformal symmetry demands that the same two-point functions are proportional to the Hessian of the $a$-function at its extremum \cite{Anselmi:1997ys}\cite{Intriligator:2003jj} . It therefore means that if we have a zero eigenvalue in the Hessian, the model cannot be realized in unitary quantum field theories.

In the above example, the moduli direction corresponds to the zero eigenvalue of the Hessian. More generally, if the $a$-function has a flat moduli direction, then the current two-point functions are always degenerate, which can be realized only in non-unitary theories. Therefore, there is no unitary superconformal field theory in which $c$ changes along the exactly marginal deformations.

We therefore conclude that since the natural mechanism to allow exactly marginal deformations cannot allow $c$ to change in unitary theories, the models in which $c$ change along exactly marginal deformations do not exist unless we resort to some sort of fine-tunings that do not rely on the supersymmetry.

\section{Holographic argument}

\subsection{Pro}

{\bf Statement:} 

{\it We do present a model in which $c$ changes along exactly marginal deformations within the effective AdS/CFT correspondence. We allow certain fine-tunings of the effective potential, but it is theoretically allowed with the controlling parameters of $1/N$. This gives us a concrete example of conformal field theories with varying $c$ under exactly marginal deformations.}

We admit that realizing exactly marginal deformations without a symmetry requires fine-tunings. This leads to a possibility of controlling the renormalization group with an extra parameter, and one way is to use the large $N$ limit by regarding $1/N$ as such. Then one may be able to obtain exactly marginal deformations without invoking quantum corrections, and one can legitimately set up the question if we can construct conformal field theories in which $c$ changes along exactly marginal deformations.\footnote{Indeed, within perturbative quantum field theories, the non-trivial corrections to $c$ with respect to gauge/Yukawa/scalar couplings have been discussed in \cite{Osborn:2016bev}\cite{Jack:1990eb}, and if the beta functions for these coupling constants vanish, then generically $c$ changes along the exactly marginal deformations. We would like to thank H.~Osborn for the contribution to the debate. See appendix A for a review of the field theory analysis in pertrurbation theory.} 
Furthermore, large $N$ theories may have a holographic description under the AdS/CFT correspondence, and we will pursue this possibility below.

As a model of the AdS/CFT correspondence, we study an effective higher derivative gravity coupled with a ``massless" scalar field $\phi$ given by the bulk action
\begin{align}
S = \int d^{5} x \sqrt{|G|} \left(\frac{1}{\kappa^2} R - \Lambda +\frac{1}{2\kappa^2} \partial^\mu \phi \partial_\mu\phi + \phi(\alpha R^2 + \beta  R_{\mu\nu}R^{\mu\nu} + \gamma R_{\mu\nu\rho\sigma} R^{\mu\nu\rho\sigma}) \right) \ .
\end{align}
As we will show, in order to assure that $\phi$ is a moduli field, we should demand $10\alpha + 2\beta + \gamma =0$.\footnote{A different combination of the higher derivative gravity that does generate a potential for $\phi$ was analyzed in \cite{Myers:2010tj}.}

We introduce the Fefferman-Graham expansions of the metric and the scalar field
\begin{align}
ds^2 &= G_{\mu\nu} dx^\mu dx^\nu = \frac{l^2}{4}\frac{d\rho^2}{\rho^2} + \frac{g_{ij} dx^i dx^j}{\rho} \cr
g_{ij} &= g_{(0)ij} + \rho g_{(1)ij} + \rho^2g_{(2)ij} + \cdots \cr
\phi & = \phi_{(0)} + \rho \phi_{(1)} + \rho^2 \phi_{(2)} + \cdots
\end{align}
and put the boundary at $\rho = \epsilon$, which plays a role of UV cut-off in the dual conformal field theory. Our goal is to compute the holographic Weyl anomaly \cite{Henningson:1998gx} from this model.

By using the boundary counterterms to cancel the UV power divergence, we may only focus on the logarithmic terms with respect to $\epsilon$ in the action: $S \sim S_{\log} \log \epsilon$.  They may be further expanded in powers of $\phi$:
\begin{align}
S_{\log} = -\frac{1}{2}  \int d^4x \sqrt{|g_{(0)}|} &\left( L_0 + \phi_{(0)} L_{00} + \phi_{(1)} L_{1} + \phi_{(2)} L_2   +\frac{2}{l\kappa^2} \phi_{(1)}^2 +\frac{l}{\kappa^2}g^{ij}_{(0)} \partial_i \phi_{(0)} \partial_j \phi_{(1)} \right. \cr
&\left.  -\frac{l}{2\kappa^2} g^{ik}_{(0)}g^{jl}_{(0)} g_{(1)kl} \partial_i\phi_{(0)} \partial_j \phi_{(0)} +\frac{l}{4\kappa^2} g^{kl}_{(0)} g_{(1)kl} g^{ij}_{(0)} \partial_i\phi_{(0)} \partial_j \phi_{(0)}  \right)\ .
\end{align}
 Here 
\begin{align}
L_0 &= \left(-\frac{6}{l\kappa^2} -\frac{l\Lambda}{2} \right) g_{(0)}^{ij} g_{(2)ij} -\left(\frac{l}{\kappa^2}\right) g_{(1)ij}R^{ij}_{(0)} + \left(\frac{l}{2\kappa^2} \right) R_{(0)}g^{ij}_{(0)}g_{(1)ij}  \cr
 &+  \left(\frac{2}{l\kappa^2} + \frac{l\Lambda}{4} \right) g^{ij}_{(0)}g^{kl}_{(0)}g_{(1)ik}g_{(1)jl} + \left(-\frac{1}{2l\kappa^2}-\frac{l\Lambda}{8} \right) (g^{ij}_{(0)}g_{(1)ij})^2 
\end{align}
and 
\begin{align}
L_{00} &=l(\alpha R^2_{(0)} + \beta R^{ij}_{(0)} R_{(0)ij} + \gamma R^{ijkl}_{(0)} R_{(0)ijkl}) \cr
&+ \left(\frac{40\alpha }{l^3} + \frac{8\beta }{l^3} + \frac{4\gamma}{l^3} \right)  g_{(0)}^{ij} g_{(2)ij} +\left( \frac{40\alpha}{l} + \frac{12\beta}{l}+\frac{12\gamma}{l} \right) g_{(1)ij}R^{ij}_{(0)} \cr
&+ \left(-\frac{8\alpha}{l}-\frac{2\beta}{l}-\frac{2\gamma}{l} \right) R_{(0)}g^{ij}_{(0)}g_{(1)ij} + \left(\frac{20\alpha}{l^3}+\frac{8\beta}{l^3} +\frac{10\gamma}{l^3} \right) g^{ij}_{(0)}g^{kl}_{(0)}g_{(1)ik}g_{(1)jl} \cr
&+ \left(\frac{6\alpha}{l^3}+\frac{2\beta}{l^3}+\frac{\gamma}{l^3} \right) (g^{ij}_{(0)}g_{(1)ij})^2 
\end{align}
as studied in \cite{Nojiri:1999mh}.

As mentioned above, we have to assure that $\phi$ is a moduli field. For this purpose, we notice 
\begin{align}
L_{2} = \alpha \frac{400}{l} + \beta \frac{80}{l} + \gamma \frac{40}{l} \ ,
\end{align}
and see that $\phi_{(2)}$ appears only in this term. Requiring that the variation of $\phi_{(2)}$ vanish then demands $10\alpha + 2\beta + \gamma =0$ as advocated. 
It is equivalent to the statement that there is no potential term for the scalar field $\phi$ in the AdS space-time. Otherwise, the assumption that $\phi$ is a moduli field does not hold and the corresponding coupling constant runs along the renormalization group flow due to the non-zero beta function.

After setting $10\alpha + 2\beta + \gamma =0$ the variation under $g_{(2)ij}$ determines $l$ from $\Lambda$ ($<0$) as
\begin{align}
-\frac{12\Lambda}{\kappa^2} = l^2 \ ,
\end{align}
which is not modified by the value of the moduli field $\phi$. As we will see, this determines the value of the Weyl anomaly coefficient $a$, and it corresponds to the fact that $a$ does not change under the exactly marginal deformations.

Finally, we have
\begin{align}
L_1 = \left( -\frac{40\alpha }{l} -\frac{8\beta }{l}-\frac{4\gamma}{l}\right) R_{(0)} + \left(-\frac{40\alpha}{l} -\frac{8\beta}{l}-\frac{4\gamma}{l} \right) g^{ij}_{(0)}g_{(1)ij} . \label{L1} 
\end{align}
Remarkably, $L_1$ vanishes once we set $10\alpha + 2\beta + \gamma =0$. There seems to exist a typo in eq (27) of \cite{Nojiri:1999mh} so that our expression is different in the second term of \eqref{L1} but after correcting it, we have this property, which is important for the consistency of the holographic Weyl anomaly as we will see.

Our next task is to derive the equations of motion to determine $\phi_{(1)}$ and $g_{(1)ij}$ in terms of $\phi_{(0)}$ and $g_{(0)ij}$ and evaluate the on-shell action.  The calculation is greatly simplified because $L_1=0$. Because of the absence of the additional mixing from $L_1$, the variation of $\phi_1$ essentially gives the same expression as studied in \cite{Nojiri:1999mh}.\footnote{If we did not correct the typo mentioned above, the expression we would get has an independent $R^2$ term which is inconsistent as mentioned in section 2.2.} Substituting the solution back into the on-shell action, we obtain the holographic Weyl anomaly\footnote{There are several different (and sometimes more convenient but eventually equivalent) ways to derive the holographic Weyl anomaly in particular in higher derivative gravities. See e.g. \cite{Fukuma:2001uf}\cite{Sen:2012fc}\cite{Rajagopal:2015lpa}\cite{Bugini:2016nvn}.}
\begin{align}
T^i_{\ i} = -a \mathrm{Euler} + c(\lambda) \mathrm{Weyl}^2 + a \left(\frac{1}{2}(\Box \lambda)^2 + R^{ij} \partial_i \lambda \partial_j \lambda -\frac{1}{3} R g^{ij}\partial_i \lambda \partial_j \lambda + \frac{2}{3}(g^{ij}\partial_i \lambda \partial_j \lambda )^2\right) \ . 
\end{align}
where
\begin{align}
a &= \frac{l^3}{8\kappa^2} \cr
c(\lambda) &= \frac{l^3}{8\kappa^2} - l \frac{10\alpha  +2\beta - \gamma}{2}\lambda \ 
\end{align}
with the condition $10\alpha + 2\beta + \gamma =0$. Here $\lambda$ is an exactly marginal deformation dual to the bulk field $\phi$.

This solution corresponds to dual conformal field theories with an exactly marginal deformation  under which the Weyl anomaly coefficient $c$ changes. As in the field theory analysis, we see that the Weyl anomaly coefficient $a$ cannot depend on the exactly marginal deformations but $c$ may. 

Therefore, we conclude that as long as we may be able to fine-tune the potential of the moduli field $\phi$, we are able to construct a higher derivative holographic model in which $c$ changes along the exactly marginal deformations.

\subsection{Con}

{\bf Statement:} 

{\it Not all effective theories of gravity are consistent. If we look at the higher derivative supergravity, again the unitarity demands that one cannot construct a model in which the holographic $c$ changes along the exactly marginal deformations.}

The supersymmetric generalization of the $O(R^2)$ term with the field dependent coupling constant was first discussed in \cite{Hanaki:2006pj}
\begin{align}
S = \int d^5x \sqrt{|G|} & \left(R - \frac{N_{IJ}}{4}F_{\mu\nu}^IF^{J\mu\nu} +  N_{IJ}\partial_\mu \phi^I \partial^\mu \phi^J  \right. \cr
&+  \left.  \frac{\epsilon^{\mu\nu\rho\sigma\tau}}{16} c_IA^I_{\mu} R_{\nu\rho}^{\ \  \alpha\beta}R_{\sigma\tau\alpha\beta}  + c_{IJK}\epsilon^{\mu\nu\rho\sigma\tau} A^I_\mu \partial_\nu A^J_\rho  \partial_\sigma A_\tau^K +\right. \cr
& \left. + c_I \phi^I \left(\frac{1}{6}R^2-\frac{4}{3}R_{\mu\nu}R^{\mu\nu} + R_{\mu\nu\rho\sigma}R^{\mu\nu\rho\sigma} \right) + V(\phi) \right) + \cdots \label{hsugra}
\end{align}
Here $\phi^I$ is a vector multiplet scalar (whose partner is $A^I_\mu$), and the supergravity model is parameterized by Chern-Simons couplings $c_{IJK}$, $c_I$ and the gauging parameter $P^I$. 

The potential $V(\phi)$ consists of two terms. First, we have the D-term constraint
\begin{align}
c_{IJK} \phi^I \phi^J \phi^K = 1 \ , \label{vs}
\end{align}
corresponding to the very special structure.
We also have the potential terms from the gauging that dictates we have to minimize
\begin{align}
P = P_I \phi^I
\end{align}
under the constraint \eqref{vs}. 

Suppose we have a supersymmetric solution with the AdS vacuum, then the potential term gives a condition that we have to minimize $ P = P_I\phi^I$ under the constraint
\begin{align}
c_{IJK} \phi^I \phi^I \phi^K = 1 \ , \label{gconst}
\end{align}
which is equivalent to the $a$-maximization condition \cite{Intriligator:2003jj}\cite{Hanaki:2006pj}. To see this, we define $t^I = \phi^I/P$, and then it is equivalent to maximize $a = c_{IJK} t^I t^J t^K$ under the condition $ P_I t^I = 1$. 
In order to find a situation in which $c$ changes along the exactly marginal deformations, we first demand that the potential has a flat direction $\frac{\partial V}{\partial \phi^I}=0$ and the corresponding Chern-Simons coupling $c_I$ is non-zero. Then, as discussed in the previous subsection, we can change $c$ along the flat moduli direction while $a$ is fixed. It is important to recognize that the combination of the higher derivative terms that appear in \eqref{hsugra} satisfies the condition $10\alpha + 2\beta + \gamma =0$ discussed in section 3.1.

From the kinematical consideration alone, it is not impossible to fine-tune $c_{IJK}$ and $P_I$ so that the supergravity action has a flat direction and holographic $c$ function changes along such a direction with non-zero $c_I$ in that direction.  However, as in the field theory analysis in section 2.2, a further requirement of unitarity makes it unphysical. The supersymmetry dictates that the coefficient $c_{IJK}$ not only determines the potential, but also the kinetic term. 
The gauge kinetic term is determined \cite{Hanaki:2006pj} from $N_{IJ} = -c_{IJK}\phi^K$ and if we have a flat direction in the potential, the gauge kinetic function is degenerate and unitarity is violated. Obviously this is the gravity dual description of the unitarity obstruction discussed in section 2.2.

Finally, when we actually have a flat direction in the moduli space of a unitary superconformal field theories, then they belong to a hypermultiplet \cite{Tachikawa:2005tq} and the above analysis does not apply. We do not know any higher derivative coupling between the hypermultiplet and $R^2$ terms in the supergravity, The field theory argument of the $a$ maximization and the t'Hooft anomaly consideration presumably makes it impossible.

We should further notice that even without supersymmetry, the naive model discussed in section 2.2 has an issue of unitarity. The unitarity demands that the central charge $c$ must be bounded below. The positivity of the two-point function demands $c>0$. 
Moreover further bound on the conformal collider physics \cite{Hofman:2008ar} (see also its proof \cite{Hofman:2016awc} from the light-cone conformal bootstrap) demands $c> \frac{18}{31}a$. If we take the naive model discussed in section 2, for larger values of $|\phi|$ one may violate this bound. To avoid this issue one has to introduce a certain cutoff for the range of possible $\phi$.\footnote{Within the effective holography, this is not difficult. One may just introduce the non-trivial field dependent kinetic term for the scalar field $\phi$.} 

We, therefore, conclude that in the supergravity in which we may control the potential better  as in some string compactifications, it is not possible to construct a model with varying $c$ along the exactly marginal directions without violating the unitarity.



\section{Debates}
The above discussions reveal that indeed we may be able to change $c$ under exactly marginal deformations in fine-tuned effective holographic models. Note that the fine-tuning here is not for the change of $c$; rather it is for obtaining exactly marginal deformations. Once this fine-tuning is made, the totalitarian principle of theoretical physics ``Everything not forbidden is compulsory" is effective and varying $c$  along exactly marginal deformations is demonstrated within the effective holography with higher derivative corrections.
Here, to conclude the paper, we are going to address two further questions: (1) was it trivial to construct the model with the desired property once we allow fine-tunings in the effective AdS/CFT correspondence? (2) what was the prospect of fine-tunings beyond the effective holography?

About the first point, we would like to emphasize that it is a non-trivial question to ask even within the effective AdS/CFT correspondence if we can actually realize various terms in the Weyl anomaly once they are allowed by the Wess-Zumino consistency condition. For example, at the conformal fixed point, the Pontryagin density in the Weyl anomaly
\begin{align}
T^{i}_{\ i} = \epsilon^{abcd}R_{abij}R^{ij}_{\ \ cd}
\end{align}
has never been realized either in holography or field theories \cite{Nakayama:2012gu}\cite{Bonora:2014qla} even though it satisfies the Wess-Zumino consistency condition.\footnote{The discussions here assume the conformal invariance. If the theory is only scale invariant without conformal invariance, we may construct such terms. See e.g. \cite{Nakayama:2013is} for a review.} There are other terms like
\begin{align}
T^i_{\ i} = C_{IJK} \epsilon^{ijk} \partial_i \lambda^I \partial_j \lambda^J \partial_k \lambda^K
\end{align}
in three dimensions \cite{Nakayama:2013wda}\cite{Kikuchi:2016ljr} or
\begin{align}
T^i_{\ i} = B_{IJ} \epsilon^{ij} \partial_i \lambda^I \partial_j \lambda^J \label{wa}
\end{align}
in two-dimensions that were not realized in holography as well as in field theories even though they satisfy the Wess-Zumino consistency conditions.

With this respect, it is non-trivial for us to be able to construct a model in which $c$ changes along exactly marginal deformations in particular because the leading Einstein gravity cannot afford it. 
Conversely it is interesting to see if there is field theory analysis that is not captured by the Wess-Zumino consistency conditions that do not allow the peculiar Weyl anomaly like the above. Indeed, in a recent paper \cite{Gomis:2015yaa}, they pointed out an obstruction to construct the Weyl anomaly \eqref{wa} from the viewpoint of the (ultra)locality in correlation functions. The argument there does not rely on the unitarity, so the nature of the varying $c$ under exactly marginal deformations is quite different from that.

The second question is related to the (in)stability of the AdS/CFT correspondence without supersymmetry. There has been a conjecture that the non-supersymmetric AdS/CFT correspondence cannot exist \cite{Ooguri:2016pdq} with various examples studied in the literature \cite{Dymarsky:2005nc}\cite{Dymarsky:2005uh}\cite{Horowitz:2007pr}\cite{Martin:2008pf}\cite{Narayan:2010em}. Our holographic construction is based on the effective AdS/CFT correspondence without supersymmetry, and if this conjecture were true, we would not have any concrete example of conformal field theories with varying $c$ under exactly marginal deformations. Alternatively, if we may prove that such conformal field theories are forbidden from the field theory argument, it gives a strong constraint on the possible AdS/CFT correspondence.

\section*{Acknowledgements}
The author would like to thank Z.~Komargodski for his kind hospitality and discussions during the author's visit to Weizmann Institute where this work was initiated.
The author would like to thank S.~Nojiri for conversions on the typo in \cite{Nojiri:1999mh}. He would like to thank Y. Tachikawa for explanations on his papers about the higher derivative supergravity.

\appendix

\section{Leading order field theory analysis}
For illustrative purposes, let us consider the $SU(N_c)$ Yang-Mills theory coupled with $N_f$ Dirac fermions in the fundamental representations. Perturbative computation at the one-loop order gives the beta function for the gauge coupling constant
\begin{align}
\beta_g = \frac{dg}{d\log\mu} = - \frac{g^3}{48\pi^2} (11N_c-2N_f)  + \mathcal{O}(g^3)
\end{align}
One may choose the matter contents so that the leading order beta function vanishes, which makes the gauge coupling constant ``moduli-like" in the one-loop approximation.

With the same order, the Weyl anomaly coefficients $a$ and $c$ have been computed (see e.g. \cite{Jack:1990eb} and references therein):
\begin{align}
a &= \frac{62}{90(8\pi)^2} (N_c^2-1) + \frac{11}{90(8\pi)^2} N_f N_c + \mathcal{O}(g^4)  \cr
c &= \frac{12}{30(8\pi)^2}(N_c^2-1) + \frac{6}{30(8\pi)^2} N_f N_c -\frac{16}{9 (8\pi^2)}(N_c^2-1)(N_c-\frac{7}{16}N_f) \frac{g^2}{16\pi^2} + \mathcal{O}(g^4)   \ .
\end{align}
This perturbative computation is in agreement with the general argument in the main text: along the (exactly) marginal deformations (i.e. $g$ in this case with $11N_c-2N_f = 0$), $a$ does not change, but $c$ may change. Note that vanishing of the leading order beta function $11N_c-2N_f=0$ does not make the leading order correction to $c$ vanish, which is proportional to $N_c-\frac{7}{16}N_f$.\footnote{In the supersymmeric case, there are non-trivial additional corrections from Yukawa and scalar couplings.}
 In this example, we do not have a control over the higher order corrections in the beta function, but we may imagine a more elaborate example in which it may be done and can be thought of as a field theory realization of the holographic model discussed in the main text.

\end{document}